# Ionic conductivity optimization of composite polymer electrolytes through filler particle chemical modification

*Andres Villa, Juan Carlos Verduzco, Joseph A. Libera[1] and Ernesto E. Marinero\**

School of Materials Engineering, Purdue University, Neil Armstrong Hall of Engineering, 701 West Stadium Avenue, West Lafayette, IN 47907.

[1] Argonne National Laboratory, 9700 South Cass Avenue, Bld. 370, Lemont, IL 60439.

*corresponding author: eemarinero@purdue.edu

**ABSTRACT**. The addition of filler particles to polymer electrolytes is known to increment their ionic conductivity (IC). A detailed understanding of how the interactions between the constituent materials are responsible for the enhancement, remains to be developed. A significant contribution is ascribed to an increment of the polymer amorphous fraction, induced by the fillers, resulting in the formation of higher ionic conductivity channels in the polymer matrix. However, the dependence of IC on the particle weight load and its composition on the polymer morphology is not fully understood. This work investigates Li ion transport in composite polymer electrolytes (CPE) comprising Bi-doped LLZO particles embedded in PEO: LiTFSI matrixes. We find that the IC optimizes for very low particle weight loads (5 – 10%) and that both its magnitude and the load required, strongly depend on the garnet particle composition. Based on structural characterization results and electrochemical impedance spectroscopy, a mechanism is proposed to explain these findings. It is suggested that the Li-molar content in the garnet particle controls its interactions with the polymer matrix, resulting at the optimum loads reported, in the formation of high ionic conductivity channels. We propose that *filler particle chemical manipulation of the polymer morphology* is a promising avenue for the further development of composite polymer electrolytes.

# DECLARATIONS

## FUNDING

This research was funded in part by Dr. Marinero's startup funds from Purdue University School of Materials Engineering and in part by the Consejo Nacional de Ciencia y Tecnologia, Mexico. (CONACYT).

## CONFLICT OF INTEREST

The authors declare that there are no conflicts of interest related to this article.

## AVAILABILITY OF DATA AND MATERIALS

The datasets generated during and/or analyzed during the current study are available from the corresponding author on reasonable request.

## CODE AVAILABLITY

Not applicable.

## AUTHORS' CONTRIBUTIONS

Not applicable.

## ASSOCIATED CONTENT

See Supplemental information

**INTRODUCTION**

Liquid electrolytes currently employed in lithium-ion batteries (LIBs) are flammable organic materials; concerns about their safety hinder the wider implementation of LIBs in critical sectors such as transportation[1]. Solid state electrolytes (SSE) with high ionic conductivity (IC) across the operating temperatures of electric vehicles are of utmost interest in current battery research. Solid polymer electrolytes (SPE) offer several advantages over ceramic materials, such as flexibility and high mechanical strength, however, they exhibit inadequate low ionic conductivities at room temperature[3].

Composite polymer electrolytes (CPE) are promising hybrid materials that could potentially circumvent the IC limitations of SPEs[2] while providing a solution to the safety issues of LIBs. Addition of $Li_7La_3Zr_2O_{12}$ (LLZO) garnet particles to PEO:salt matrixes[4–6] is reported to increase the polymer matrix ionic conductivity. The increase in ionic conductivity is attributed, amongst other, to a decrease of the extent of crystallinity in the polymer matrix upon filler addition[7, 8]. The material properties of the fillers, including particle size[9], surface chemistry[7], and intrinsic ionic conductivity[10] have also been reported to strongly influence ionic transport.

The cubic phase of LLZO was first reported by Murugan et al.[11] to exhibit ionic conductivities as high as $3 \times 10^{-4}$ S/cm at 25°C and following their work, significant efforts to further increment IC, have focused on aliovalent substitution at the Li, La and Zr sites and on implementing sintering processes to achieve the highest ceramic pellet densification. For example, Dong et al.[12] synthesized $Li_{6.6}La_3Zr_{1.6}Ta_{0.4}O_{12}$ garnets and utilized high-pressure spark plasma sintering at 1050°C to fabricate pellets with relative densities of 0.94 that exhibited an ionic conductivity of $1.18 \times 10^{-3}$ S/cm at room temperature.

In addition to the nature of the filler material, the amount of filler necessary to optimize IC has been found to depend on the ratio of EO:Li provided by the solvated Li-ionic salts and on the polymer molecular weight (MW). This amount has been reported for different filler materials to range from ~5%wt. to ~52% wt.[5, 6, 10, 13].

Utilizing relatively low contents of filler materials[5, 14] the ionic conductivity of the polymer-salt matrix can be drastically improved. Potentially, further improvements in constituent material properties of CPEs could result in ionic conductivities matching or exceeding that of pure garnet ceramic electrolytes. This would decrease the amount of rare-earth (such as La) materials needed, thereby lowering the electrolyte material cost. In addition, unlike ceramic garnet electrolytes, CPEs require no energy intensive and costly sintering processing to form high density solids, thereby drastically reducing the manufacturing cost of solid-state electrolytes. The polymer molecular structure plays an important role in determining IC in CPEs and some semi-crystalline systems have been reported to outperform fully amorphous structures[15], in addition, several reports indicate that CPEs with the lowest degree crystallinity of the polymer do not provide the highest IC [16–18].

Recently, further increments of CPE ionic conductivity have been demonstrated by engineering the particle morphology and their alignment to generate within the polymer matrix fast ion transport channels. These efforts include optimization of the particle shape[19, 20] and methods to achieve particle alignment within the polymer matrix[20, 21]. Additional approaches have explored the utilization of dual-layer polymer electrolytes, wherein a PEO-based layer is used as an anolyte/separator structure and another layer serves as a catholyte/cathode-binder [22]. In-situ polymerization and additives have been employed also to improve the formation of interphases between layers[23]. Therefore, it is evident, as pointed out in a recent review on the status of PEO+LLZO composite electrolytes[24], that combining engineered CPE microstructures designed for high ionic transport with constituent material property improvements, offers a wide range of options to enhance the performance of CPEs and therefore, an attractive opportunity to realize and implement a low-cost manufacturing option for the fabrication of SSEs[24].

In this work we investigate ionic transport in CPEs comprising PEO: LiTFSI matrixes embedded with garnet particles having different Bi-content, namely $Li_6La_3ZrBiO_{12}$ (Bi-LLZO), $Li_{6.25}La_3Zr_{1.25}Bi_{0.75}O_{12}$ (0.75Bi-LLZO) and $Li_{6.25}La_{2.8}Nd_{0.2}Zr_{1.25}Bi_{0.75}O_{12}$ (0.75BiNd-LLZO). We find that the Bi content of the garnet particles determines the percentage weight (%wt.) load required to achieve the highest ionic conductivity of the composite materials studied. Bi-aliovalent substitution at the Zr-site modifies the garnet particle Li-molar content to maintain charge neutrality and in turn, the magnitude of the IC of the CPEs studied. To distinguish between effects due to lattice parameter changes, brought about by partial substitution of Zr by Bi, vs. changes in the Li-site vacancy occupancy, La was partially substituted by Nd while keeping constant the Li-molar content. IC measurements of CPEs loaded with these dual-doped garnet particles indicate that the Li-molar content is the dominant factor. To investigate whether the observations here reported are associated to the CPE microstructure, we utilized characterization techniques (XRD, SEM, DSC, FTIR, Polarized Light Microscopy) in conjunction with electrochemical impedance microscopy to correlate morphological changes of the polymer matrix induced by the garnet particles to the CPE ionic conductivity. Based on analysis the data, we propose a mechanism that ascribes the reported results to the Li-molar content of the LLZO particles. We suggest that the Li-vacancy occupancy determines the particle physico-chemical surface properties and thereby, its interactions with the polymer:ionic salt matrix and in particular, the polymer morphology.

This results in the formation of interconnected high conductivity amorphous channels and both, the optimum particle weight load and the magnitude of the CPE ionic conductivity are controlled by the nominal Li-molar content in LLZO.

**MATERIALS AND METHODS**

$Li_{7-x}La_{3-y}Nd_yZr_{2-x}Bi_xO_{12}$ garnet powders were synthesized using a sol-gel Pechini method described in prior publications from our group[25–28]. To compensate for Li loss during calcination, an excess of 10% of $LiNO_3$ (99.0%, Sigma Aldrich) was utilized. Bi doping was chosen for aliovalent substitution in LLZO garnets at the Zr-site, as it stabilizes the cubic phase formation at significantly lower calcination temperatures than when other dopants are added[25–29]. The high-IC garnet cubic phase was obtained upon powder calcination at 700°C for 10h. The Bi-valence state on the of $Zr^{4+}$ site is reported[29] to be $Bi^{5+}$. Synchrotron EXAFS studies reveal that Bi occupies the Zr-site exclusively[27]. Bi doping modifies the garnet stoichiometry and to retain charge neutrality, the Li-molar content is proportionally changed[29]. Bi additions to LLZO also improves densification upon sintering, a key requirement for attaining the highest IC in pure garnet ceramic solids[30].

Composite polymer electrolytes were fabricated by mixing in acetonitrile (ACN, Sigma Aldrich, 99.9%) calcined cubic phase garnet powders with PEO (MW=100,000 Sigma Aldrich,) and lithium bis(trifluoromethanesulfonyl)imide (LiTFSI, Sigma Aldrich, 99.8%). ACN was mixed at a 2.5:1 liquid to solids ratio to form the PCE slurry. The materials were wet ball-milled for 12h at 400 rpm with a Fritsch Pulverisette 6 apparatus. This yielded garnet particles with average size of ~437nm ($d_{50}$). An EO:Li ratio of 49:1 was selected, based on prior studies in our group[26] that determined the optimum ratio required to maximize IC in PEO (MW=100,000) - Bi-doped LLZO composite solid membranes. To study the effect of the filler particle intrinsic ionic conductivity on the IC of PEO:LiTFSI matrixes, we compared additions of Bi-LLZO garnet particles to those obtained when loaded with non-ionically conductive $Al_2O_3$ particles of comparable size ($d_{50}$=500nm, Almatis, 16-SG). All films were dried for 72h at room temperature in air to allow slow evaporation of ACN. Thereafter, the films were held in vacuum overnight to completely remove the solvent. Additional synthesis details are provided in the supplementary information.

The formation of cubic phase LLZO garnet powders, following calcination at 700°C, was verified by XRD. Synthesis parameters were adjusted based on XRD measurements to obtain the pure cubic phase and to suppress secondary phases and trace impurities. A Bruker D-8 Focus apparatus equipped with a Cu source (1.54Å) was employed and patterns were acquired with a scan rate of 5°/min. ImageJ software[31] analysis of SEM images of wet-ball milled powders was used to determine average LLZO particle size. Details of the particle size determination are provided in the supplementary information. The particle dispersion in the composites was checked using SEM analysis and the degree of PEO crystallinity in the CPE membranes was investigated using XRD. A scan rate of 2.5°/min was employed for these measurements. (for additional information on particle dispersion, refer to the supplementary information). The polymer morphology evolution as a function of LLZO particle additions was examined utilizing polarized light microscopy (PLM). Differential scanning calorimetry (DSC) was employed to determine the melting points of CPEs loaded with different LLZO particle amounts, as well as variations in polymer crystallinity. We note that the stoichiometries of the sol-gel synthesized garnet particles given in this work are nominal values and are controlled by accurate weighing of the precursor materials. Utilizing the same sol-gel procedure for the synthesis of Al-LLZO, Ca-LLZO, Ba-LLZO and Sr-LLZO garnet materials, others have shown excellent agreement (with differences ranging from 0 to 2.5%) between nominal stoichiometry and of that obtained, utilizing inductively coupled plasma spectroscopy (ICP)[32, 33].

To verify that the synthesis procedure used in this work to vary the Bi content of LLZO results in changes of the Li-molar content of the garnet, the stoichiometry of garnet powders with various nominal compositions were measured employing ICP-OES (Inductively Coupled Plasma-Optical Emission Spectroscopy) at Argonne National Laboratory. Ceramic powders were fabricated using the same sol-gel synthesis procedure outlined in this work. The compositional results together with a description of the measurements are reported in the supplementary information. Measured ICP molar contents for Bi, Zr and Li are presented in Table S2 and a correlation study between measured and nominal values, which are controlled by the chemical concentrations of precursor materials in the sol-gel synthesis, is presented in Fig. S8. The agreement between the Bi and Zr nominal and measured values are in excellent agreement, whereas, the measured Li-molar content, appears to be higher than the nominal values. The measured Li-molar content for all samples is ~0.37 (±0.07) higher than that expected from the measured Bi-content. This is likely to originate from the excess $LiNO_3$ that is used to compensate for Li volatilization during calcination of the powders. Nevertheless, these preliminary measurements do confirm an inverse relationship between the Li-molar content and increasing Bi-doping of the LLZO garnets. We also note that CPE fabrication involves dissolution of the garnet powders in ACN which is known to remove Li-carbonates formed at the garnet particle surface. The measured powders were not dissolved in ACN and therefore, they may contain a higher amount surface Li-carbonate, which may be reflected in the higher Li content measured by ICP-OES. Precise determination of Li-content in electrode materials by ICP-OES is challenging and requires multiple measurements using large sample volumes. We plan to conduct in future experiments a more exhaustive determination of the Li-molar content in Bi-LLZO garnet materials, in particular when incorporated into the polymer matrix.

Ionic conductivity was characterized using electrochemical impedance spectroscopy (EIS). Temperature-dependent measurements were performed at 22.5°C, 35°C, 45°C, and 55°C, with a custom-built Swagelok-type apparatus using two stainless steel electrodes whose separation was adjusted to optimize their contact with the membranes. Reliable measurements required full contact between the electrode surfaces and the polymer membranes, as well as attaining temperature stability. A period of 2h at the target temperature was introduced prior to measurements in order to ensure temperature stability and equilibration between the samples and the measurement electrode apparatus. A Solartron SI 1260 impedance/gain-phase analyzer and SI 1287 electrochemical interface operating in the frequency range from 100 mHz to 300 kHz was employed. An AC potential of 50 mV was applied to the membranes. Representative examples of Nyquist plots are presented in the supplementary information (figure S5) together with a description of the methodology employed to extract IC values.

**RESULTS AND DISCUSSION**

Fig. 1 presents XRD scans for the Bi-LLZO, 0.75Bi-LLZO and 0.75BiNd-LLZO calcined powder samples. Fig. 1(d) corresponds to the reference pattern for cubic LLZO (ICSD collection code 185540). As seen in the figure, all samples exhibit the garnet cubic phase, and no evidence of the tetragonal phase is observed. The peak at 28.45° is due to trace amounts of $La_2Zr_2O_7$, a common byproduct in LLZO synthesis. Rietveld analysis, as detailed in the supplementary information, was utilized to determine the amounts of $La_2Zr_2O_7$ present in Bi-LLZO and 0.75Bi-LLZO powder samples, namely ~2.08%wt. and ~1.03%wt., respectively.

The temperature dependence of the IC of polymer composites embedded with garnet particles of different nominal Li-molar content and particle %wt. loads are presented in Fig. 2. To quantify the effect on ionic conductivity induced by the garnet particle additions to the polymer matrix, the results obtained for the PEO:LiTFSI matrix alone and for a matrix loaded with 5%wt. non-ionically conducting $Al_2O_3$ particles are presented also in Figs. 2b) - 2d). The influence on IC of the Li-molar content of the garnet particles is shown in Fig. 2a) in which the IC for CPEs loaded with 5%wt. of 0.5Bi-LLZO, 0.75Bi-LLZO and 1.0Bi-LLZO particles are compared over the temperature range 22.5°C to 55°C. The composite loaded with 1.0Bi-LLZO garnet particles, the lowest Li-molar content, exhibits the highest IC. The influence of the Li-molar content on the IC of CPEs, to our knowledge, has not been priorly reported and it is quite remarkable, in particular given the wide separation of the embedded particles in the polymer matrix, which precludes intra-particle ion transport as the dominant conductivity mechanisms. To further study the effect of %wt. load on the temperature dependence of IC, we compare samples loaded with 0.75Bi-LLZO and Bi-LLZO. We find that the optimum %wt. load depends also on the garnet Li-molar content. This is shown in Figs. 2b) and 2c) in which the temperature dependence of the IC for samples loaded with Bi-LLZO (Fig. 2b) and 0.75Bi-LLZO (Fig. 2c) are presented. It is noted that at 22.5°C and 45°C, addition of 5% wt. load of Bi-LLZO increments the IC of the PEO-LiTSI matrix by factors of ~22.5x and ~67x respectively. Whereas for additions of 0.75Bi-LLZO particles, the highest IC is obtained when the particle wt. load is 10%; the corresponding IC increments determined at 22.5°C and 45°C are ~9x and ~36x, respectively. The values of IC for the CPE loaded with 5% wt. of Bi-LLZO particles at 22.5°C and 45°C were measured as $2 \times 10^{-5}$ S/cm and $4.7 \times 10^{-4}$ S/cm, respectively. For comparison, the corresponding values at 22.5°C and 45°C for the filler-free PEO-LiTSI matrix were measured as $8.9 \times 10^{-7}$ S/cm and $7.2 \times 10^{-6}$ S/cm, respectively.

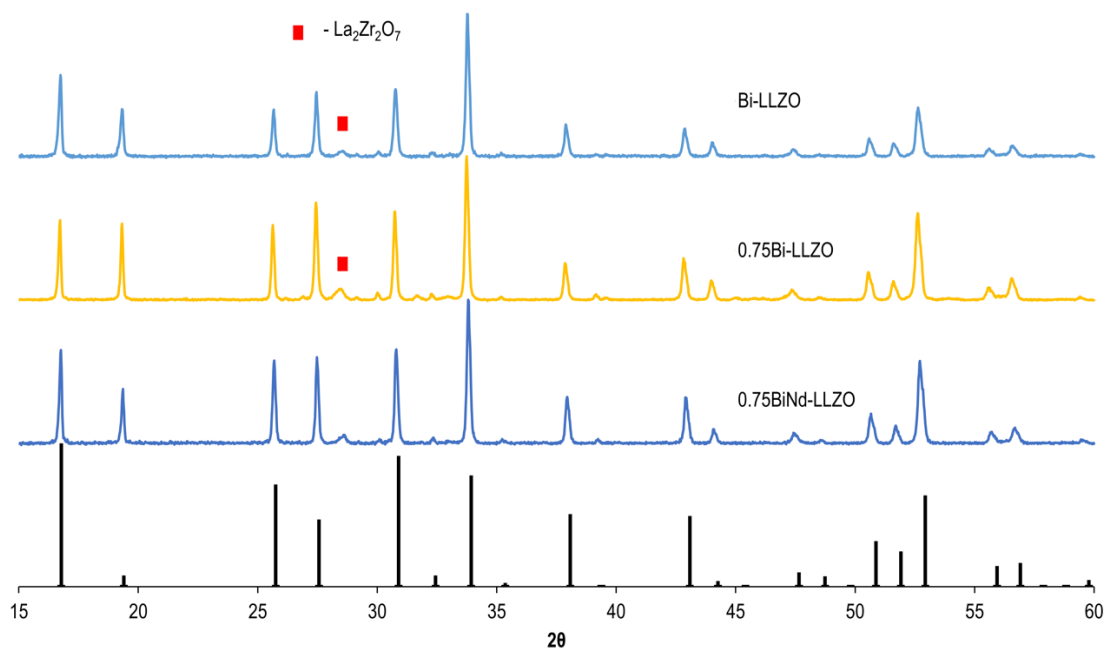

**Figure 1.** XRD patterns corresponding to calcined garnet powders of this study: (a) 0.75Bi-LLZO; (b) Bi-LLZO; (c) 0.75BiNd-LLZO, (d) reference PDF pattern corresponding to the cubic phase of LLZO (ICSD collection code 185540).

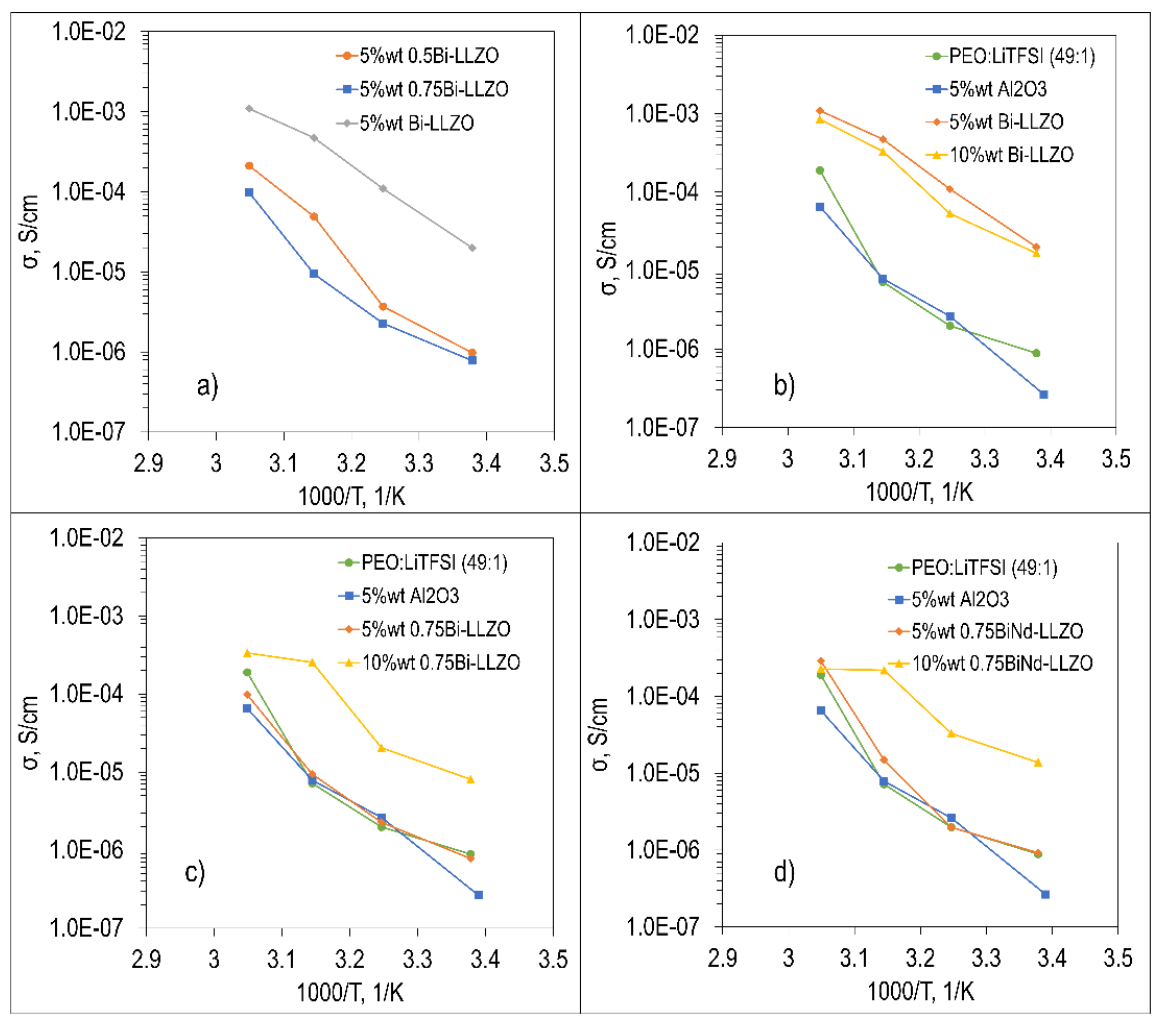

**Figure 2.** Temperature dependence of IC for CPEs loaded with **a)** 5%wt Bi-doped LLZO and Al$_2$O$_3$ particles, the nominal Bi content of the garnets is 0.5, 0.75 and 1.0; The effect of changing the particle %wt. load from 5% to 10% is shown in Figs. b) – d); **b)** Bi-LLZO particles; **c)** 0.75Bi-LLZO; **d)** 0.75BiNd-LLZO particles. For comparison, the IC of the PEO-LiTFSI matrix and that of the matrix loaded with 5% Al$_2$O$_3$ particles is also presented in these figures.

To further investigate whether the effects observed are correlated to the garnet Li-molar content rather than to the type of dopant or to lattice distortions brought about by changing the Bi-content, we synthesized garnets with dual dopants at different lattice sites, while maintaining the Li-molar content fixed. To this effect we doped LLZO with both Bi and Nd to form the garnet with the nominal composition Li$_{6.25}$La$_{2.8}$Nd$_{0.2}$Zr$_{1.25}$Bi$_{0.75}$O$_{12}$ (0.75BiNd-LLZO). It is noted that Nd exclusively substitutes on the La-site and Bi on the Zr site. Fig. 2d) presents the temperature and particle %wt. load dependence of PEO:LiTSI matrixes embedded with 0.75BiNd-LLZO. Remarkably, the same behavior is observed as for the case of additions of 0.75Bi-LLZO particles (Fig. 2c): IC maximizes for the same wt. load (10%) and values at 22.5°C and 45°C are essentially the same. This indicates that the garnet Li-molar content plays a key role on IC of the CPEs here studied.

The temperature dependence of the IC of CPEs loaded with 10% wt. of 0.75Bi-LLZO and 0.75BiNd-LLZO particles (Figs. 2c) and 2d)) indicates that the IC conductivity at the highest measured temperature (55°C) tend to a saturation value. To investigate whether an Arrhenius behavior is observed in these composites, the data for these materials is plotted as $Ln(\sigma T)$ vs. 1000/T, the results are provided in Fig. S6 of the SI together with a discussion of the analysis.

The results shown in Fig. 2, consistent with published results, indicate that the addition of garnet particles increments the IC of the PEO-LiTFSI matrix. On the other, *and no priorly reported*, we find that the load amounts needed to maximize the CPE ionic conductivity, depend on the LLZO particle composition and more specifically, on the Li-molar content. In addition, the highest IC values are observed for samples embedded with garnet particles having the lowest Li-molar content. As mentioned earlier, to maintain charge neutrality in $Li_{7-x}La_3Zr_{2-x}Bi_xO_{12}$, the Bi-substitution into the Zr-site results in modifications of the Li content. Hence, as Nd is incorporated into the La-site, the Li content when co-doping Nd and Bi into LLZO is controlled by the Bi amount. These observations indicate the importance of the garnet particle Li content.

Fig. 3 presents the IC dependence of the composites vs. %wt. load at room temperature and 45°C for Bi-LLZO, 0.75Bi-LLZO and 0.75BiNd-LLZO particle additions. Fig. 3a) compares the %wt. load dependence for Bi-LLZO and 0.75Bi-LLZO, whereas in Fig. 3b) that for 0.75Bi-LLZO and 0.75BiNd-LLZO is provided. The results show that embedding garnet particles with different Bi-content requires different %wt. loads for optimum IC. Furthermore, it is evident that the highest IC values are obtained at both temperatures in composites embedded with Bi-LLZO particles (lowest Li-molar content). Fig. 3b) compares the incorporation into PEO-LiTFSI matrixes of 0.75Bi-LLZO and 0.75BiNd-LLZO particles on IC. The results are virtually identical, regarding both the %wt. load required, and the actual magnitude of the IC measured at both temperatures. A decrease in IC is observed in all samples studied at higher particle loads than the optimum value, and at 50%wt. load, the IC approaches the value measured in non-loaded PEO-LiTFSI matrixes.

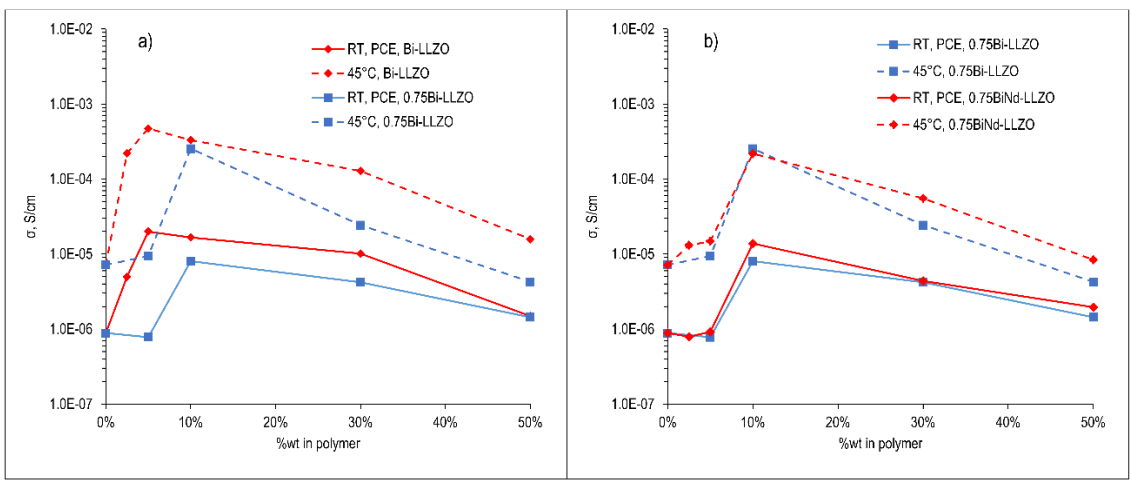

**Figure 3.** Ionic conductivity at 22.5°C and 45°C of PEO:LiFTSI samples vs. particle %wt. load for a) Bi-LLZO and 0.75Bi-LLZO; b) comparison of IC of CPEs loaded with 0.75Bi-LLZO and 0.75BiNd-LLZO garnet particles.

To determine whether the IC dependence on particle %wt. load and Li-molar content reported in this work can be ascribed to significant changes in the amorphous fraction of the polymer matrix, XRD was used to characterize the composites. Fig. 4 presents XRD patterns of composites wt. loaded with 5% Bi-LLZO, 10% 0.75Bi-LLZO and 0.75BiNd-LLZO particles (these loads are the optimum values to achieve the highest IC in each composite). The measured XRD patterns are compared in Fig. 4 to those corresponding to the PEO:LiTFSI matrix alone and to the composite loaded with 5%wt. $Al_2O_3$ particles. For reference, the XRD pattern corresponding to the LLZO cubic phase is also included in the figure. The diffraction peaks in the 19°- 23.5° range correspond to crystalline PEO and decrements in their intensity with particle additions indicate a reduction in polymer crystallinity. The scans show for the case of the samples loaded with garnet particles, peaks corresponding to cubic LLZO and their intensities become stronger as the %wt. of filler particles is incremented. Whereas a decrease of the intensity of the peak at 23.5° is observed with the addition of the 5%wt. Bi-LLZO, 10%wt. 0.75Bi-LLZO and 10%wt. 0.75BiNd-LLZO, no change is measured for the case of 5%wt. $Al_2O_3$ additions.

This suggests that garnet particle additions to PEO:LiTFSI increment the polymer amorphous fraction. However, we find that larger garnet particle additions, while further reducing this peak intensity, implying formation of higher amounts of amorphous fraction, rapidly decreases IC beyond the optimum value. Thus, the results indicate that at the optimum particle loads required to provide the highest IC in the composites, no drastic crystalline to amorphous transformation occurs and that the magnitude of the polymer amorphous fraction alone, does not explain why the IC maximizes at the low particle %wt. loads reported in this work. This is consistent with published results on CPEs that also report that the maximum IC does not necessarily correspond to the lowest degree of crystallinity in the polymer phase[16–18].

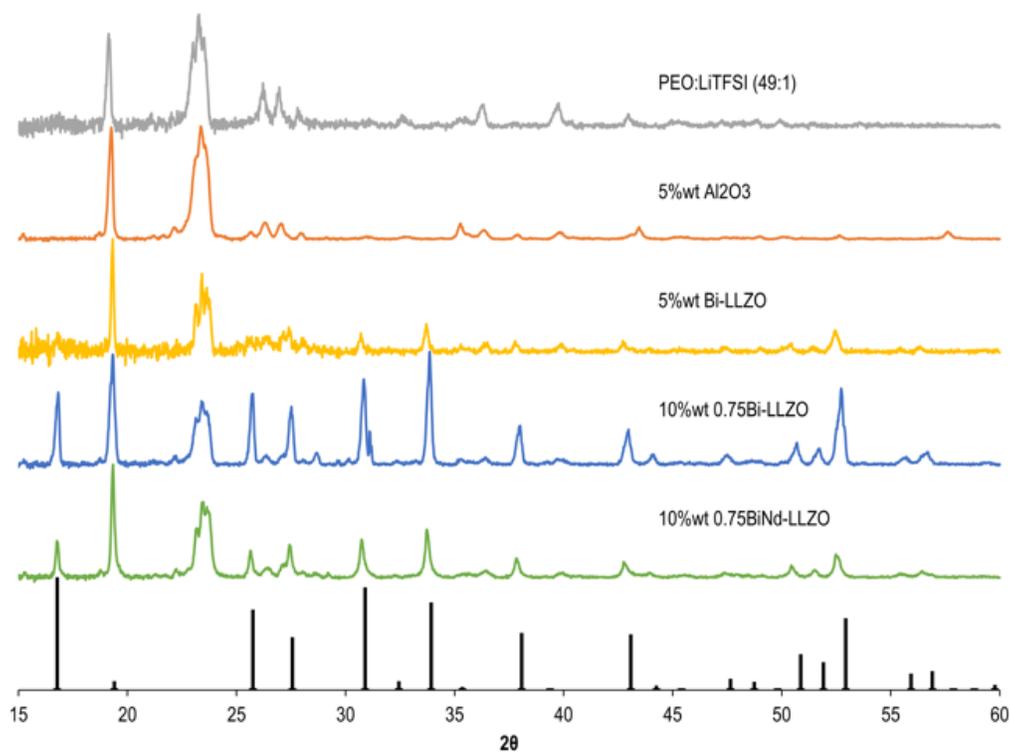

**Figure 4.** XRD patterns for PEO:LiTFSI matrixes loaded with the particle %wt. loads of 0.75Bi-LLZO, Bi-LLZO and 0.75BiNd-LLZO that yield the highest IC in their composites. For comparison the patterns for the unloaded PEO:LiTFSI matrix and that loaded with 5%wt. $Al_2O_3$ particle are also shown. For reference, the PDF pattern corresponding to cubic LLZO (ICSD collection code 185540) is included. The diffraction peaks in the 19°- 23.5° range correspond to crystalline PEO and decrements in their intensity are interpreted to imply a reduction in polymer crystallinity.

To further investigate structural changes of the polymer matrix induced by the addition of garnet particles, DSC measurements were conducted to determine melting temperatures and fusion enthalpies as a function of particle wt. load amounts. DSC curves for PEO:LiTFSI films loaded with increasing Bi-LLZO particle amounts are presented in Fig. S4 of the supplementary information. Melting points were determined from endothermic peaks occurring between 55°C and 60°C upon sample heating. The melting temperatures for all CPEs studied is ~ 57°C, with only a small decrease observed from 57.9°C for the unloaded polymer matrix to 57.3°C for that loaded with 5%wt. Bi-LLZO particles. Similarly, the enthalpy of fusion, decreased slightly from 107.3 J/g for the unloaded sample, to 103.5 J/g for the sample loaded with 5%wt. Bi-LLZO particles. Consistent with the XRD results, the DSC measurements indicate that no drastic crystalline to amorphous transformation is associated with the optimum garnet particle %wt. load required to attain the highest IC in the composites.

To investigate other polymer morphological changes that may explain the garnet particle composition and wt. load dependence of IC, polarized light microscopy (PLM) was used to study spherulite formation at and near the optimum loads needed to yield the highest IC values. Fig. 5 provides PLM images acquired at 35°C of composite membranes loaded with 5% and 10%wt. loads of Bi-LLZO, 0.75Bi-LLZO and 0.75BiNd-LLZO particles. PLM measurement details are provided in the supplementary information. Higher magnification images were utilized to resolve spherulite boundaries to estimate their size average. The following spherulite average diameter changes were observed when incrementing the particle wt. load in the polymer matrix from 5% to 10%: ~98μm to ~39μm for Bi-LLZO; ~87μm to ~119μm for 0.75Bi-LLZO and 58μm to ~214μm for 0.75BiNd-LLZO. For these particular samples, the associated IC changes measured were from $1.09 \times 10^{-4}$ S/cm to $5.33 \times 10^{-5}$ S/cm for Bi-LLZO; from $2.27 \times 10^{-6}$ S/cm to $2.05 \times 10^{-5}$ and from $9.2 \times 10^{-7}$ S/cm to $1.38 \times 10^{-5}$ S/cm for 0.75BiNd-LLZO. The changes in spherulite size revealed by PLM indicate that their size is influenced by both the garnet particle %wt. load and its Li molar content. The spherulite average size measurements indicate that incrementing the Bi-LLZO wt. load from 5% to 10% reduces their average size; whereas the opposite occurs for the case of 0.75Bi-LLZO and 0.75BiNd-LLZO particle additions. Furthermore, the highest IC is obtained when Bi-LLZO particles are added to the CPE. For the case of 0.75Bi-LLZO and 0.75BiNd-LLZO particle loadings, the highest IC is obtained at the same particle wt. load (10%) and its magnitude is approximately the same for both samples. The influence of spherulite formation on polymer ionic conductivity has been reported by Marzantowicz *et al.*[34] in their study of spherulites in PEO:LiTFSI as a function of the EO:Li ratio (50:1 and 6:1). They concluded that the drastic decrease of conductivity for the EO:Li=6:1 sample is related to the closing of amorphous conductivity pathways by growing spherulites. Likewise, Xi *et al.*[35, 36] utilizing PLM and DSC investigated the effect of incorporating ZSM-5 molecular sieves on the crystallization kinetics of PEO and showed that adding fillers to PEO influences the heterogeneous nucleation and growth of spherulites in the polymer matrix.

The results here reported provide new findings on these hybrid materials, specifically that the *chemical composition of the garnet filler particles and in particular, their Li-molar content,* influences spherulite formation and growth. We find that by controlling the garnet Bi-doping and consequently, the Li-molar content, the particle %wt. load can be optimized as well as the CPE ionic conductivity. It is suggested that the physico-chemical and surface properties of the garnet particles influence the nucleation and growth of spherulites and the evolution of the polymer morphology that is conducive to facile ion transport.

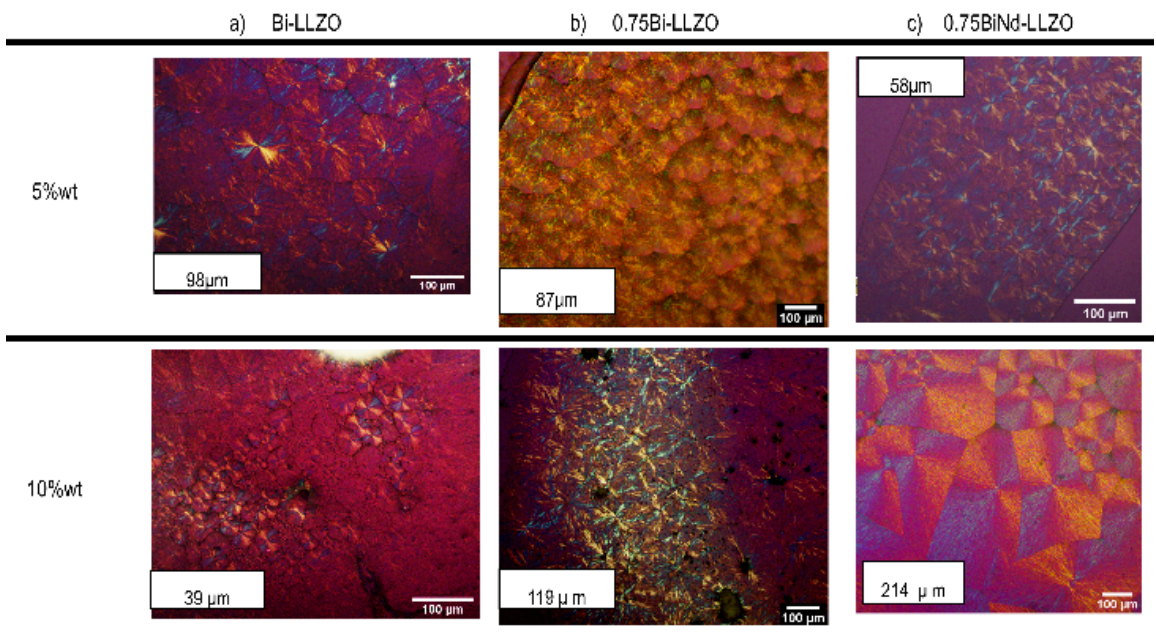

**Figure 5** Polarized light microscopy images of spherulites in PEO:LiTFSI polymer matrixes with 5% and 10%wt. loads of a) Bi-LLZO; b) 0.75Bi-LLZO and c) 0.75BiNd-LLZO. Images were taken at 35°C. Changes in spherulite average size are clearly observed with particle %wt. load changes. See supplementary information for details.

To understand the relationship between spherulite formation and ionic conductivity, it is noted that spherulites typically grow radially from nucleation centers and their ultimate size is controlled by growth kinetics and volume competition. When filler particles are introduced into polymers, spherulites nucleate both from random defect sites in the polymer and heterogeneously from the filler particle surface. The results of Fig. 5 indicate that in our case, heterogeneous nucleation at garnet particle surfaces dominate spherulite formation and growth. Spherulites comprise crystalline lamellae separated by higher ionic conductivity amorphous regions. At the boundaries between spherulites, interconnection between amorphous regions can facilitate long range transport, whereas connections between crystalline and amorphous zones, hinder transport.

The most favorable situation for long-range ion transport arises when large spherulites are formed and their amorphous regions interconnect at the boundaries, thus providing macroscale high conductivity transport channels in the polymer matrix. This description helps to understand the IC dependence on garnet particle %wt. load: for very small particle amounts, the low density of heterogenous nucleation centers results in the formation of isolated spherulites with little or no connectivity. Conversely, when the particle load is too high, a large number of spherulites are nucleated and volume competition for growth, results in the formation of numerous small spherulites as well as a multiplicity of spherulite boundaries. This gives rise to different interconnections between amorphous/amorphous, crystalline/crystalline and amorphous/crystalline boundaries. Ionic conductivity is thus, hindered due to an increment in tortuosity of the amorphous channels and the formation of resistive boundaries at crystalline-to-amorphous interfaces. The optimum IC ensues when the %wt. load of filler particles yields a critical density of nucleation centers conducive to the formation of large spherulites that interconnect through a minimum number of amorphous to amorphous regions at the boundaries. This results in the formation of long range interconnected amorphous channels, thereby forming macroscale high ionic-conductivity polymer pathways. This description is consistent with observations by Cheng et al.[37] and Fullerton et al.[38] regarding ion transport in polymer electrolytes. *A key observation from our work is that the filler particle chemistry controls the nucleation and growth of spherulites* in composite polymer:ionic salt matrixes doped with garnet particles.

We note that the creation of high ionic conductivity channels in CPEs has been pursued by using filler particles to modify the polymer morphology, for example by optimizing the particle shape[19, 20] or by aligning the particles[21, 39, 40] to generate high ionic transport channels in the bulk of the polymer.

Significant improvements in IC have been achieved using these approaches; however, these methods add complexity to the fabrication of solid-state electrolytes. In contrast, the mechanism here proposed, *polymer morphology manipulation by controlling the filler particle chemistry,* is simpler and takes advantage of the interactions between the filler particle surface and chemical properties and the polymer matrix to form high ionic conductivity channels

.

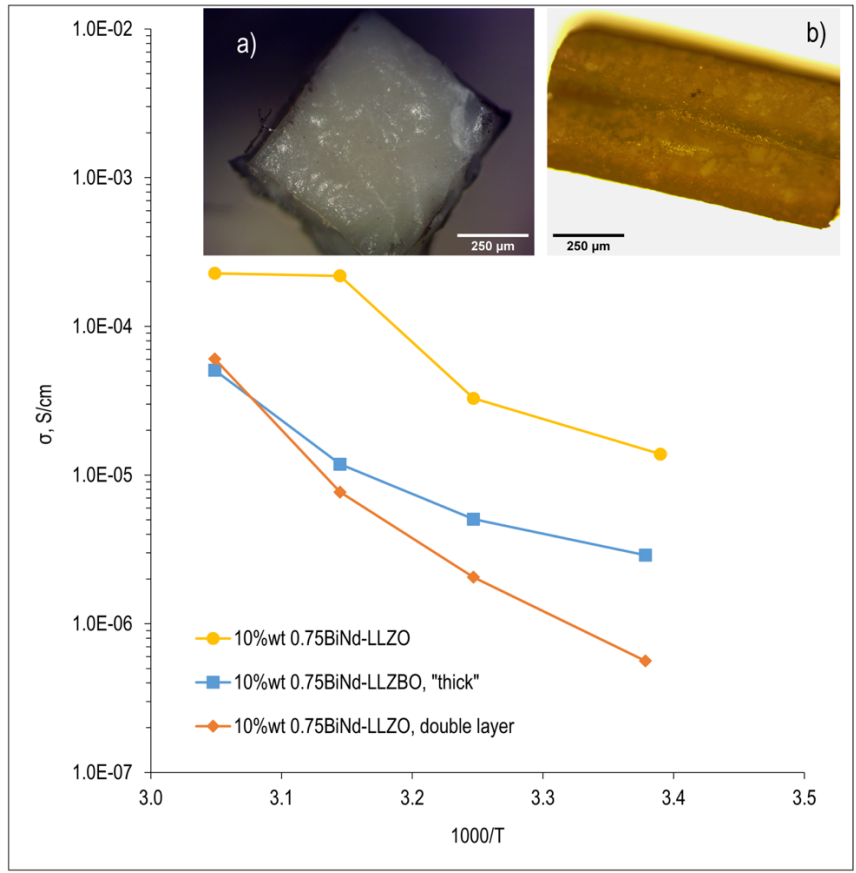

**Figure 6.** Ionic conductivity vs. temperature for CPE membranes of different thicknesses loaded with 10%wt. 0.75BiNd-LLZO particles. The thickness of the membranes corresponding to the plot labelled 10%wt. 0.75BiN-LLZO is 310 μm; *thick* refers to a ~710μm sample, and *double layer* is made of two separate membranes bonded by annealing; the total thickness of this sample is ~500μm. Inset images correspond to a) thick film and b) double layer bonded film.

Finally, to provide additional support on the role of the polymer morphology on ionic conductivity, we investigated its dependence on membrane thickness and fabrication. Increasing the thickness, augments the spherulite volumetric density and the number of boundary zone interconnections along the transport direction. Likewise, bonding together two thin CPE membranes, makes it statistically unlikely to join the amorphous regions from each thin layer component. Therefore, we expect transport to be negatively impacted in both situations.

The typical thickness of the CPE membranes studied in this work is ~310μm and the average spherulite size measured is ~119μm in polymer matrixes loaded with 10%wt. 0.75BiNd-LLZO particles. Assuming spherical spherulite shape, it implies that along the ion transport direction in the EIS measurements, there is a maximum of at most, three interconnected spherulites in these ~310μm thick membranes. As expected, and presented in Fig. 6, increasing membrane thickness or bonding two thin membranes together, decremented IC in CPEs loaded with 10%wt. 0.75BiNd-LLZO particles. The highest IC across the temperature range studied corresponds to the 310μm membrane. The thicker membrane (~710μm) and the bonded dual layers (~500μm thick) exhibit lower ionic conductivity. An important corollary of this experiment is that thinner membranes should significantly enhance the ionic conductivity in these hybrid materials. Early results from our work support this conclusion and pave the way to further increment IC in these composite solid-state electrolytes.

**CONCLUSIONS**

In this work, the effect of adding Bi-doped LLZO particles to PEO:LiTFSI polymer matrixes to form composite polymer Electrolytes (CPE) has been investigated. The addition of garnet particles increased the ionic conductivity of the PEO:LiTFSI matrices by factors ranging from ~10x to ~67x across the temperature range studied (22°C to 55°C). In particular, 5%wt. additions of Bi-LLZO particles resulted in IC increments of 67x at 45°C. The enhancement in ionic conductivity is ascribed to the formation of a polymer morphology comprising a network of interconnected spherulites having an optimum size and density. A plausible model, based on structural characterization of the composites and on EIS measurements, is proposed to explain the observed effects. Bi-aliovalent substitution at the Zr-site in LLZO changes the nominal Li-molar ratio from Li=7.0 in LLZO to Li=6 in Bi-LLZO and to Li=6.25 in 0.75Bi-LLZO and 0.75BiNd-LLZO garnets respectively, thereby altering the density of Li vacancy occupancy. Compositional analysis of garnet powders by ICP-OES confirms that the Li-molar content is modified by varying the Bi-doping of the garnet particles. We suggest that changes in Li-molar content modifies the filler particle chemistry and its surface properties. The embedded particles provide heterogeneous spherulite nucleation centers in the polymer matrix and the Li-molar ratio of the garnet particles, determine in turn, the %wt. load needed to form the optimum polymer morphology that facilitates macroscopic ion transport in these composite hybrid materials.

In summary, the results presented in this work provide evidence on the critical role played by the polymer morphology of CPEs on ionic conductivity. Furthermore, our results indicate that altering Li-molar content of the garnet particle controls the formation of the desired polymer microstructure to facilitate ion transport. *Control of the* *polymer morphology by chemical manipulation of filler particles* in hybrid composite polymer electrolytes presents an attractive approach to enhance IC in this class of promising solid-state electrolytes. This study opens the possibility to further enhance ion transport in CPEs through chemical manipulation of the polymer morphology together with optimization of constituent material properties. These include the polymer molecular weight and its chemistry, the nature of the ionic salt and of the EO:Li ratio, the garnet composition, and the effect of different aliovalent dopants, as well as drastic reduction of the membrane thickness. On the practical side, the approach here demonstrated, can potentially decrease the fabrication costs of solid-state electrolytes while merging the best attributes of polymers and ceramic materials. We conclude by also noting that further studies are needed to provide a comprehensive mechanistic understanding of the interactions between the constituent materials of polymer composite electrolytes, in particular on the role of the filler particle surface chemistry and physical properties on the overall IC of these promising solid-state electrolytes.

**ACKNOWLEDGMENTS**

It is a pleasure to acknowledge the following undergraduate students at Purdue University for their help with some aspects of the experimental work and measurements: David Starr, Andee Reynolds, and Mai Tai. Fruitful discussions with Professors David Bahr and Jeffrey Youngblood at Purdue are also acknowledged.

**CONFLICT OF INTEREST**

The authors declare that there are no conflicts of interest related to this article.

# Ionic conductivity optimization of composite polymer electrolytes through filler particle chemical modification

*Andres Villa, Juan Carlos Verduzco, Joseph A. Libera[1] and Ernesto E. Marinero\*.*

\*School of Materials Engineering, Purdue University, Neil Armstrong Hall of Engineering, 701 West Stadium Avenue, West Lafayette, IN 47907.

[1] Argonne National Laboratory, 9700 South Cass Avenue, Bld. 370, Lemont, IL 60439.

**SUPPLEMENTAL INFORMATION**

This section provides additional information on sample synthesis, measurements and details on the derivation of parameters reported in the main text. To decrease the average particle size of the calcined Bi-doped garnet powders, they were mixed with polyethylene oxide (PEO, Sigma Aldrich, molecular weight 100,000), lithium bis(trifluoromethanesulfonyl)imide (LiTFSI, Sigma Aldrich, 99.8%) and acetonitrile (ACN, Sigma Aldrich, 99.9%) was used as the solvent to fully mix the composite materials. ACN was mixed at a 2.5:1 liquid to solids ratio to form a CPE slurry. All the materials were ball milled for 12 hours at 400 rpm with a Fritsch Pulverisette 6 apparatus. Milling for 12h at 400 rpm yielded mesoparticles of ~430nm ($d_{50}$) in size. High purity 100mm x 42mm x 18mm yttrium stabilized zirconia boats were obtained from MTI Corporation for high temperature calcination of ceramic precursors. A 100ml yttrium stabilized zirconium oxide jar with lid and yttrium stabilized zirconium oxide spherical grinding media (5mm in diameter) were purchased from Across International for ball milling experiments.

The choice for the milling time and rpm was based on prior studies from our group[1] of the dependence of particle size on wet-ball milling process parameters. Fig. S1 presents SEM images of powders milled at 400 rpm for 1h, 2h, 6h and 8h. The average particle size was observed to decrease from ~ 1µm to ~ 680nm. Increasing the milling time to 12h resulted in an average particle size of ~430nm with significant reduction of large particles.

Average garnet particle size was derived from SEM images of wet ball-milled samples dispersed on carbon substrates. The SEM magnification was calibrated to provide accurate size measurements using the images scale bar. SEM image analysis was performed using ImageJ software [2]. Grayscale images were processed and converted to binary images. This conversion process segments the images into particles (black contrast) or background (white contrast). As part of this conversion process, lower and upper thresholds are adjusted to include all possible particles. The degree of particle circularity was identified by assigning them values between 1.0 and 0, where 1.0 corresponds to a perfectly circular particle, whereas 0 identifies an elongated polygon. This process was repeatedly iterated to include all possible particles. Particles that deviated significantly from the chosen parameters were not included automatically in the data set; they were added by manually outlining the particle boundaries using the Wand tool. After completing the conversion process and obtaining the binary images, the *Analyze Particles* software command is employed to count and measure the particle dimensions and then compile the number and area of the particles measured. The total number of particles and associated areas are then tabulated and exported to a spread sheet. The diameter of each individual particle is then calculated, and a histogram of particle dimensions is created. The reported $d_{50}$ value in this work is defined as the median particle size derived from the analysis described.

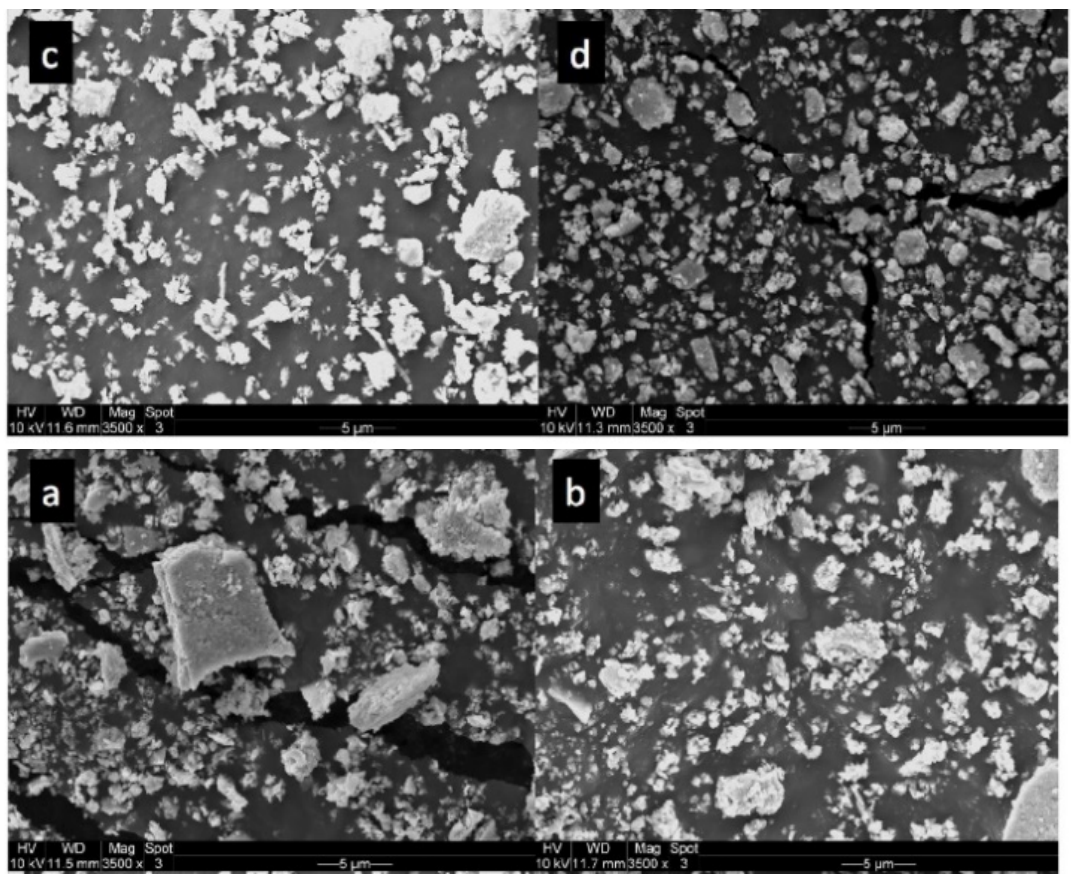

**Figure S1** SEM images of ball-milled Bi-doped LLZO garnet particles as a function of milling time: a) 1h, b) 2h, c) 4h and d) 8h. Significant particle size reduction is observed for long milling times. The average particle sizes were estimated using ImageJ analysis to decrease from ~1μm to ~ 680nm.

An example of a particle size measurement after ball milling the LLZO cubic phase powder for 8h at 400 rpm is shown in Fig. S2. A total of 358 particles were measured and the particle size range is given by the x-axis. The height of the histogram bins corresponds to the frequency of occurrence (given by the left y-axis) of particles having the same size. The data points of the orange plot (solid symbols) provide the cumulative percentage (indicated by the right y-axis) of particles having an average size of 400nm or smaller. d50 for this distribution is given by the size (x-axis) corresponding to the point where a horizontal line drawn from y= 50% intercepts the solid curve (orange plot).

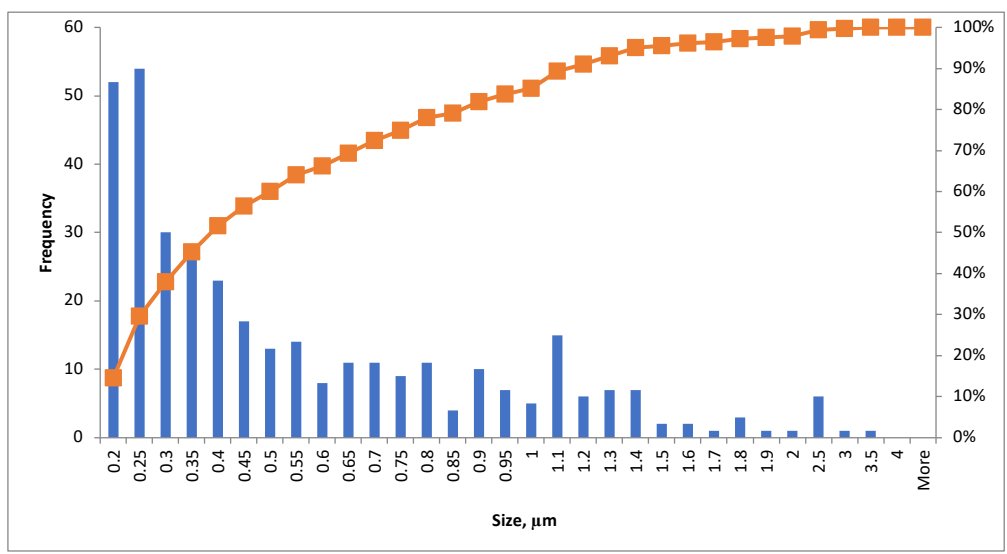

**Figure S2** Particle size distribution analysis for Bi-LLZO garnet particles after 8h of ball-milling at 400 rpm. The orange curve corresponds to the cumulative percentage of particles having a size ($d_{50)}$) ~400nm or smaller. See text for details.

To obtain a homogeneous distribution of garnet in the film, the resulting slurry after ball-milling was syringed onto 25mm x 25mm x 1mm PDMS molds. The solvent was allowed to evaporate slowly over 72h at room temperature at atmospheric pressure, with an additional 24h under vacuum at room temperature. Fig. S3 presents SEM images of PEO:LiTSFI polymer matrixes loaded with Bi-LLZO particles. Fig. S3a) corresponds to the CPE loaded with 5%wt. Bi-LLZO particles and it shows a rather uniform particle distribution within a contiguous and smooth polymer surface. Note the large separation of the garnet particles and as described in the main text, this particle load yielded the highest IC in this CPE. Loading the polymer matrix with Bi-LLZO 50%wt. as shown in Fig. S3b) resulted in a rougher surface, exhibiting cracks in the middle of the image. The IC at this particle %wt. load differs little from that of the filler-free PEO:LiTFSI matrix. We note that these images are representative of the particle distribution at the CPE surface.

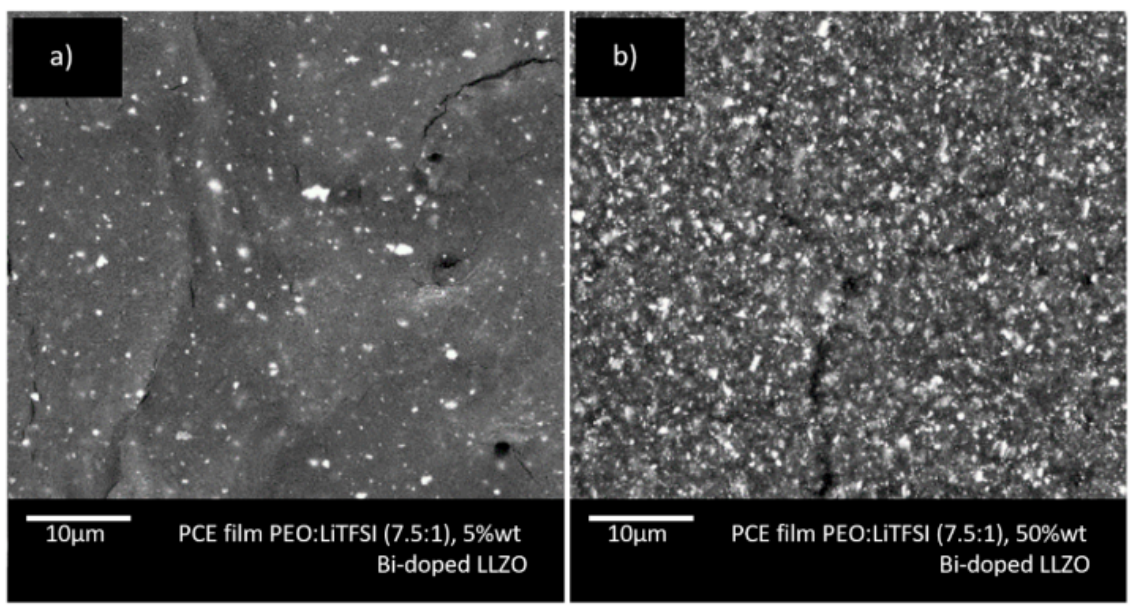

**Figure S3** SEM images of Bi-doped garnet particles dispersed in PEO:LiTFSI. The corresponding particle wt. load in a) is 5%, whereas for b) is 50%. A relative uniform dispersion of particles is observed in both cases, however, the high particle %wt. load results in the formation of cracks in the polymer matrix. Membrane thickness ~ 3.50μm

Melting temperatures and fusion enthalpies of CPE films were measured by DSC with a TA Instruments Q2000 equipment. The samples were sealed in Al pans, heated to 100°C, then cooled down to -100°C, and finally heated back up to 100°C, at a rate of 20°/min. Figure S4 shows DSC scans for the unloaded PEO:LiTFSI film, those loaded with different %wt. loads of Bi-LLZO particles, and with 5%wt. load of $Al_2O_3$ particles. The heat flow is normalized to the mass of PEO. Measured fusion enthalpies and melting points are reported in Table S1.

**Table S1** Values of $\Delta H_M$ and $T_M$ for CPE films embedded with Bi-LLZO and $Al_2O_3$ particles

| wt. Load (%) | $\Delta H_M$ (J/g) | $T_M$ (°C) |
|---|---|---|
| 0% | 107 | 57.9 |
| 5% $AL_2O_3$ | 106 | 58.6 |
| 2.5 | 103 | 57.2 |
| 5 | 104 | 57.3 |
| 10 | 103 | 57.9 |
| 30 | 98.7 | 57.2 |
| 50 | 104 | 57.9 |

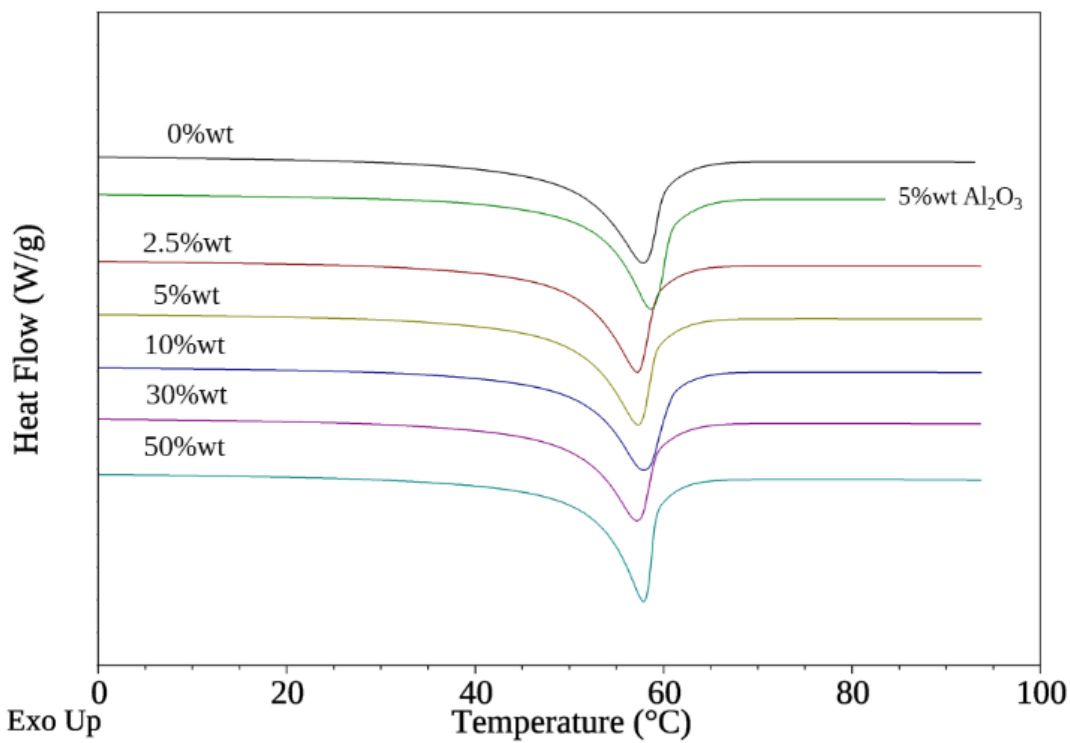

**Figure S4** DSC scans of PEO:LiTFSI membranes with different %wt. loads of Bi-LLZO particles. For comparison, the measurements for the unloaded polymer and that for a sample loaded with 5%wt. of $Al_2O_3$ particles is also displayed. The DSC scans are shifted vertically for clarity.

To investigate the effect of film thickness on the IC of CPEs, three types of films were prepared. Standard films (~310μm) were cast on a 25mm x 25mm x 1mm mold using a syringe to pour the slurry. To prepare "thick" films (<710μm), the slurry was poured with a syringe into a similar square-shaped 2mm thick casting mold. Finally, to prepare the "double" films (~500μm), standard film membranes were dabbed with a q-tip with acetonitrile (ACN) and then pressed together. Optical microscopy was used to characterize the thickness of the samples. To study spherulites utilizing PLM, the following procedure was employed to obtain suitable samples for analysis. Films were prepared using the same procedure used to perform (EIS) characterization. The samples were then loaded onto the sample fabrication mold and heated to 99°C while pressing on them with a glass slide for 10min using a 200g weight, the purpose of this annealing and pressing procedure was to decrease the sample thickness to facilitate light transmission through the sample. Upon cooling to room temperature, the top glass slide was removed, and the thinned down film was placed on the casting mold for PLM observations. First, the microscope heating stage temperature was increased to 99°C and the polymer was then cooled down from the melt at a rate of 5°C/min, and held at 35°C. We note that observations were only made in the 5%wt. and 10%wt. garnet loaded samples, as at higher particle loads, the samples were not translucent enough to acquire high quality images due to garnet particle light absorption and scattering.

We note that the described methodology for sample preparation for PLM studies has been successfully utilized in studies of spherulite formation in PEO:LiTFSI thin films[3, 4]. Although the thermal history of samples characterized by PLM in our work differs from those measured by EIS, the experimental procedure used to prepare all PLM samples was identical. This argues in favor that the trends observed regarding the spherulite growth as a function of particle %wt. load are valid. However, we recognize that the kinetics of spherulite nucleation and growth is influenced by the thermal history of the sample and therefore the spherulite dimensions reported in this work likely would differ, were the samples not first exposed to 99°C.

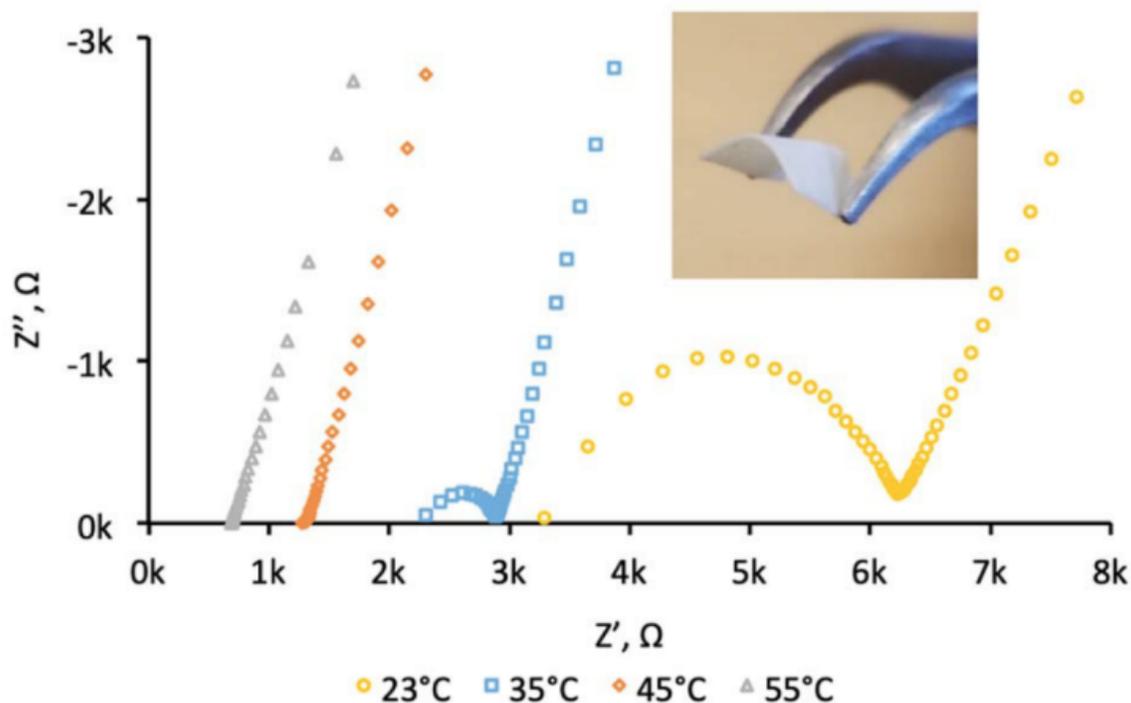

**Figure S5** Nyquist plots for Bi-LLZO polymer composite electrolyte films at 23°C, 35°C, 45°C and 55°C. Inset is an image of tweezers holding a sample CPE with 5%wt. load of Bi-LLZO particles. Note the film flexibility.

Measurements of the ionic conductivities for the composites were conducted using electrochemical impedance spectroscopy (EIS) and the measurements were analyzed and fit using an RC model with both resistive and capacitive

$$\sigma = \frac{t}{AR} \qquad (S1)$$

components in parallel, where the magnitude of the real component of the semi-circle observed, for example in figure S5 at 23°C, represent the resistive component, R. Using the measured thickness and area of the sample, equation S1 is employed to determine the IC.

As the measurement temperature is increased, the semi-circle decreased in magnitude, and started to disappear at 45°C, indicating increasing capacitive behavior and decreasing resistance for ionic conduction.

To determine whether the temperature dependence of IC of the composites obeys Arrhenius behavior, data from Fig. 2c) and 2d) is plotted in Fig. S6 as stLn(σT) vs. 1000/T. The activation energy for ion transport in such case can be estimated from equation S2, which is derived from the Nernst-Einstein equation and expresses the relationship between diffusion and ionic conductivity.

$$\sigma = \frac{\sigma_0}{T} e^{-\frac{E_A}{k_B T}} \qquad (S2)$$

Where T is absolute temperature (K), $k_B$ is Boltzmann's constant, $\sigma_o$ is the pre-exponential factor and $E_A$ is the activation energy for ion conduction.

Linear regression analysis of the data was utilized to derive estimates of the activation energies for ion transport in CPEs loaded with 10% 0.75Bi-LLZO and 0.75BiNd-LLZO particles. The linear fits yield activation energy values of 1.1 eV and 0.84 eV for 0.75Bi-LLZO and 0.75BiNd-LLZO and the fits have R-square values of 0.92 and 0.91 respectively. As all IC measurements are conducted below the melting point of our samples, determined from DSC analysis as ~57°C, one would expect an Arrhenius relationship for the temperature range studied. However, it is apparent from inspection of Figs. 2 and S6 that deviation from linearity can be observed. In particular the increment in IC between 45°C to 55°C is considerably slower than between 22.5°C to 45°C. We consider two plausible explanations for this behavior. The first possibility is that the ionic conductivity measured at 55°C is close enough to the polymer matrix melting point, so that the composite microstructure may be transitioning from a semi-crystalline solid to a more fully amorphous phase[5]. This would entail a change in the dominant ionic transport mechanism, from ion transport along the interconnected high conductivity spherulite amorphous regions proposed in this work to a mechanism facilitated by the random segmental motions of the polymer chains.

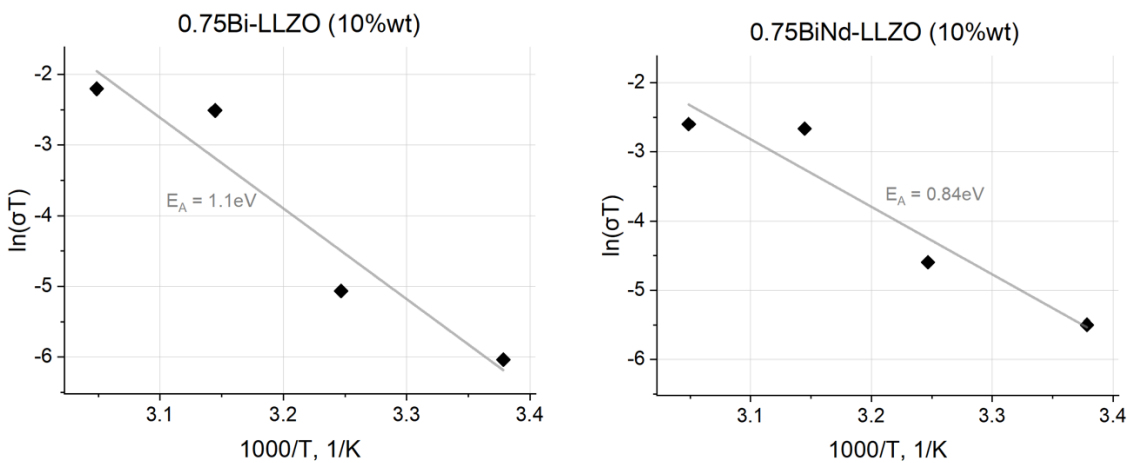

**Figure S7.** Activation energy derivation, assuming Arrhenius behavior for PEO:LiTFSI loaded with 10% wt of 0.75Bi-LLZO and 0.75BiNd-LLZO particles. The change in IC between 45°C and 55°C is slower than at lower temperatures. See text for discussion for mechanisms that could account for this change.

The second possibility is that the deviation from linearity observed is suggestive that the transport mechanism is more correctly described by the Vogel-Tamman-Fulcher (VTF) model and a better fit would be a curved line. This would imply that the composites loaded with 10%wt. 0.75Bi-LLZO or 0.75BiNd-LLZO garnet particles do not considerably change in amorphous fraction across the temperatures studied in this work [6]. We note that depending on the particle load[7] in CPEs, IC has been reported to exhibit either Arrhenius or VTF behavior. To validate which behavior is the dominant one in the CPEs here reported, further studies with a larger number of temperature measurements is needed.

XRD was used to determine the phase purity of the calcined powders before ball-milling. MAUD was used to perform Rietveld refinement of the data[11]. To determine the amount of $La_2Zr_2O_7$ [12] present in the calcined powders, the reference CIS file reported by Hamao[13] was employed and the appropriate substitution of Bi atoms at known Li and Zr sites[14]. The Li1 and Li2 site occupation information based on published work[15–17] was also was also employed. Figure S7 shows the fitting for Rietveld refinement for Bi-LLZO and 0.75Bi-LLZO. For the case of Bi-LLZO the crystalline $L_2Zr_2O_7$ phase was estimated at ~2.08%wt, while for the 0.75Bi-LLZO powders it is estimated at ~1.03%wt. These calculations were in agreement and in similar values to those reported previously, where a value of ~3%wt was reported[18].

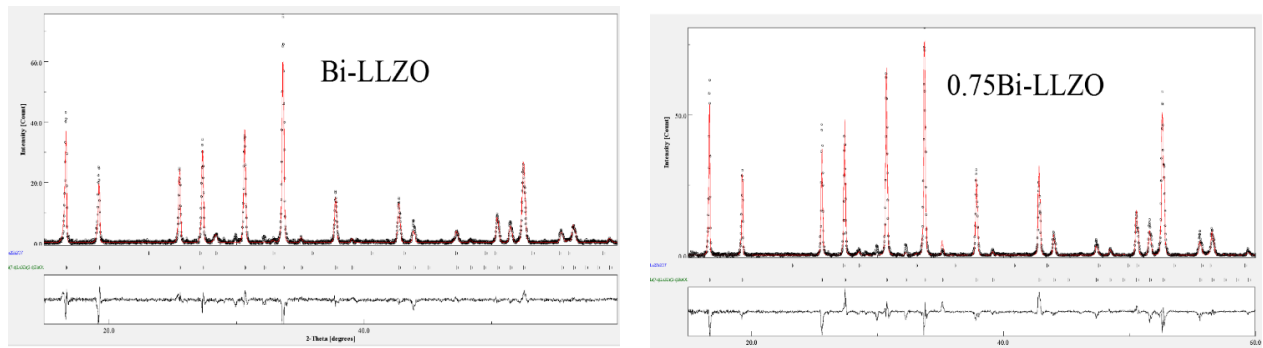

**Figure S7.** Measured (black) and calculated (red) XRD patterns of Bi-LLZO and 0.75Bi-LLZO garnets powders. The difference between measured and calculated values are shown below the measured XRD patterns. The weak peak at 28.45° indicates the presence of trace amounts of $L_2Zr_2O_7$.

ICP-OES measurements were conducted in order to determine the Li-molar content of garnet powders synthesized using the procedures describe in this work and having nominal Bi-contents of 0.5Bi-LLZO, 0.75Bi-LLZO and Bi-LLZO. For the digestion of the samples, 75 mg of each material was added to a mixture of sulfuric acid, nitric acid and 30% hydrogen peroxide and heated in a CEM Mars 6 Microwave Digestion system. Digested samples were diluted in 100 mL of deionized water. A second 1:10 dilution with 2% nitric acid was prepared and run through an Agilent ICP-OES 5110 apparatus at Argonne National Laboratories. The wavelengths employed for the elements analyzed are as follows: Al (167.019nm), Bi (223.061nm), Li (670.783nm), La (333.749nm), and Zr (339.198nm). ICP determines the mole fractions of the garnet constituent elements. The results given in Table S2 are normalized against the stoichiometric lanthanum content. We note that very small trace amounts of Al are also detected in the analysis which are likely to originate from either sample handling or contamination during powder processing.

**Table S2. Characterization of Bi content as dopant in LLZO, results for ICP-OES.**

| Composition | Li | La | Zr | Bi | Al |
| --- | --- | --- | --- | --- | --- |
| **0.5Bi-LLZO** | 6.89 | 3 | 1.53 | 0.53 | 0.01 |
| **0.75Bi-LLZO** | 6.5 | 3 | 1.27 | 0.8 | 0 |
| **Bi-LLZO** | 6.42 | 3 | 1.02 | 1.05 | 0.02 |

Linear regression analysis of the correlation between ICP-OES measured Bi, Zr and Li molar content and the nominal molar values that are controlled by the sol-gel synthesis process are presented in Fig. S8. The correlation is exceptionally good for the intended Bismuth and Zirconium substitution of the garnet material. The measured values different from the nominal ones by a scale factor that can be attributed to an error in sensitivity factor in OES analysis or to uncertainty in the concentrations of the starting chemicals in the sol-gel synthesis. The remarkable linearity bespeaks of the quality of the garnet synthesis process and the analysis for ICP measurements. In comparison, the correlation between the nominal Li-molar content and the ICP measurements deviates from perfect linearity and the regression analysis shows only a reasonable R-squared value of 0.87. The Li-content determined by ICP appears to be higher by ~0.37 (±0.07) than that expected from the measured Bi-content. This is likely to originate from the excess $LiNO_3$ that is used to compensate for Li volatilization during calcination of the powders. Determination of Li in electrode materials is challenging and measurement uncertainties of ~4% are not uncommon, precise measurements require multiple sample measurements and large sample volumes. Nevertheless, and the most important conclusion from the ICP-OES analysis, is that as shown in the lower right-hand panel of Fig. S8, incrementing the Bi-content in Bi-LLZO decreases the garnet Li-molar content.

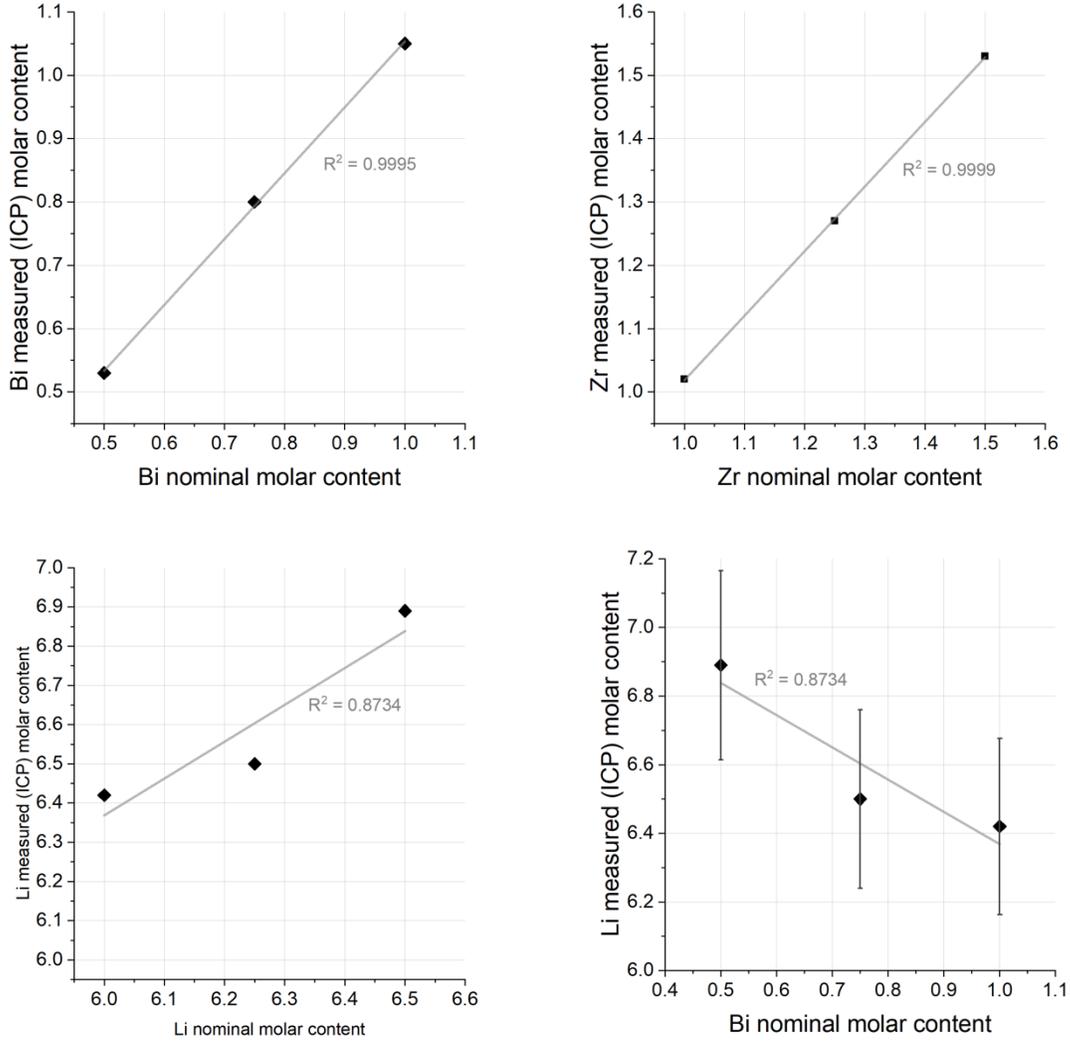

**Figure S8.** Correlation between Bi, Zr and Li molar content determined by ICP-OES and the target nominal values controlled in the sol-gel synthesis of $Li_{7-x}La_3Zr_{2-x}Bi_xO_{12}$ garnets by the concentrations of the starting chemicals. Note the exceptionally good correlation between measured and targe values for Bi and Zr. Although the linearity for the case of the Li-molar content is lower, the results clearly indicate that incrementing the Bi-content in the garnet materials, decreases the Li-molar content. See manuscript and the SI text for additional information.

SUPPLEMENTAL REFERENCES